\documentclass[sigconf]{acmart}
\AtBeginDocument{%
  \providecommand\BibTeX{{%
    \normalfont B\kern-0.5em{\scshape i\kern-0.25em b}\kern-0.8em\TeX}}}

\usepackage{booktabs} 
\usepackage{multirow}
\usepackage{latexsym}
\usepackage{graphicx} 
\usepackage{algorithm}
\usepackage{algorithmic}
\usepackage{epsfig}
\usepackage{color}
\usepackage{amsmath}
\usepackage{graphicx,subfigure}
\usepackage{makecell}
\usepackage{textcomp}
\usepackage{url}
\usepackage{flushend}
\usepackage{bbm}
\usepackage{multirow}

\usepackage{bbding}

\copyrightyear{2022}
\acmYear{2022}
\setcopyright{acmcopyright}
\acmConference[SIGIR '22] {Proceedings of the 45th International ACM SIGIR Conference on Research and Development in Information Retrieval}{July 11--15, 2022}{Madrid, Spain.}
\acmBooktitle{Proceedings of the 45th International ACM SIGIR Conference on Research and Development in Information Retrieval (SIGIR '22), July 11--15, 2022, Madrid, Spain}
\acmPrice{15.00}
\acmISBN{978-1-4503-8732-3/22/07}
\acmDOI{10.1145/XXXXXX.XXXXXX}

\begin{document}
\title[FUM: Fine-grained and Fast User Modeling for News Recommendation]{FUM: Fine-grained and Fast User Modeling\\ for News Recommendation}

\author{Tao Qi}
\affiliation{%
  \institution{Department of Electronic Engineering,\\ Tsinghua University}
  \country{}
}
\email{taoqi.qt@gmail.com}

\author{Fangzhao Wu}
\authornote{The corresponding author.}
\affiliation{%
  \institution{Microsoft Research Asia}
  \country{}
}
\email{wufangzhao@gmail.com}

\author{Chuhan Wu}
\affiliation{%
  \institution{Department of Electronic Engineering,\\ Tsinghua University}
  \country{}
}
\email{wuchuhan15@gmail.com}

\author{Yongfeng Huang}
\affiliation{%
  \institution{Department of Electronic Engineering,\\ Tsinghua University}
  \country{}
}
\email{yfhuang@tsinghua.edu.cn}

\begin{abstract}

User modeling is important for news recommendation.
Existing methods usually first encode user's clicked news into news embeddings independently and then aggregate them into user embedding.
However, the word-level interactions across different clicked news from the same user, which contain rich detailed clues to infer user interest, are ignored by these methods.
In this paper, we propose a fine-grained and fast user modeling framework (\textit{FUM}) to model user interest from fine-grained behavior interactions for news recommendation.
The core idea of \textit{FUM} is to concatenate the clicked news into a long document and transform user modeling into a document modeling task with both intra-news and inter-news word-level interactions.
Since vanilla transformer cannot efficiently handle long document, we apply an efficient transformer named Fastformer to model fine-grained behavior interactions.
Extensive experiments on two real-world datasets verify that \textit{FUM} can effectively and efficiently model user interest for news recommendation.

\end{abstract}



\maketitle

\section{Introduction}

News recommendation methods can alleviate the information overload, which are important for improving user experience and developing smart cities~\cite{konstan1997grouplens,das2007google,lin2014personalized,qi2021uni}.
A critical step of news recommendation is to accurately model the interest of a target user~\cite{zheng2018drn,qi2021kim,wutanr}.
Existing methods usually first independently encode user's clicked news into news embeddings and then aggregate them to build user embedding~\cite{wu2021personalized,qi2020privacy,qi2021hierec,yi2021efficient,qi2021pprec,danzhu2019}.
For example, \citet{wu2019neuralc} first employ the self-attention mechanism to learn news embeddings for user's clicked news from their titles, independently.
Then they attentively aggregate embeddings of clicked news to learn user embedding.
\citet{an2019neural} propose to first utilize a CNN network to learn representation of each news from their titles and categories.
They further learn user representations from embeddings of user's previously clicked news via a GRU network and ID embeddings.

\begin{figure}
    \centering
    \resizebox{0.42\textwidth}{!}{
    \includegraphics{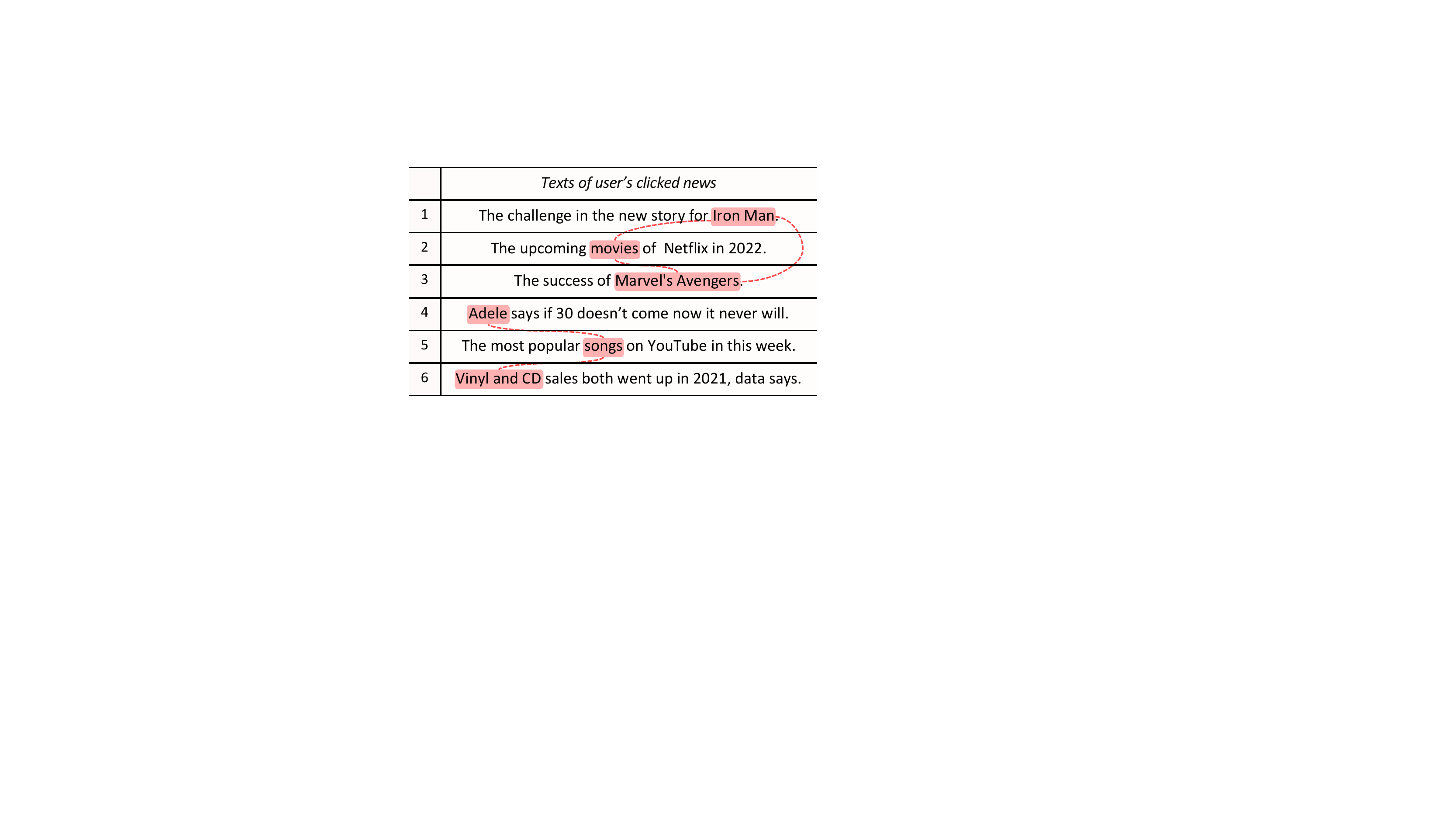}
    }
    \caption{The news clicked by a randomly selected user. Word-level relatedness across texts of user's clicked news provide detailed clues to understand user interest.}
    \label{fig.motivation}
\end{figure}


In fact, word-level interactions across clicked news from the same user contain rich detailed clues to understand user interest~\cite{wang2020fine}.
For example, according to the reading history of an example user in Fig.~\ref{fig.motivation}, we can infer the user may be interested in the movie of Iron Man from the relatedness of the word ``movies'' in the 2-nd clicked news and the word ``Iron Man'' in the 1-st clicked news.
Besides, we can also target the potential user interest in the song of Adele from the relatedness of the word ``Adele'' in the 4-th clicked news and the word ``songs'' in the 5-th clicked news.
However, most of the existing methods neglect word-level behavior interactions when modeling user interest, which may lead to inferior user modeling.



\begin{figure*}
    \centering
    \resizebox{0.95\textwidth}{!}{
    \includegraphics{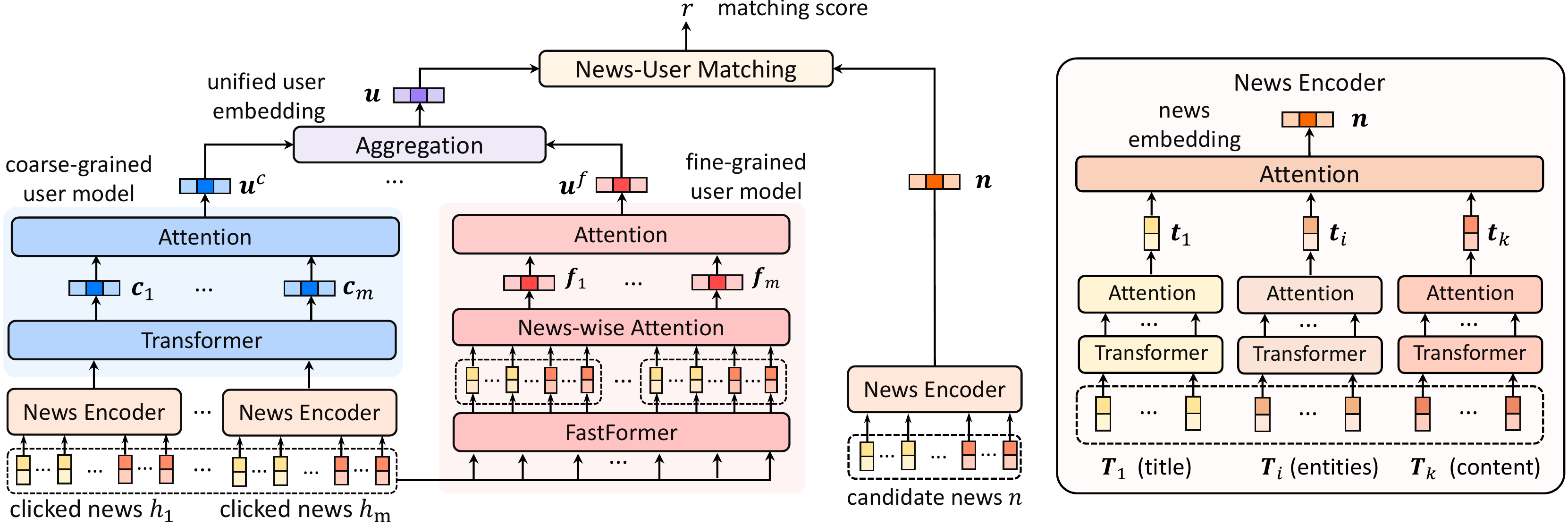}
    }
    \caption{The framework of \textit{FUM} for news recommendation.}
    \label{fig.user_model}
\end{figure*}

In this paper, we propose a fine-grained and fast user modeling framework (named \textit{FUM}) to model user interest from fine-grained behavior interactions for news recommendation.
In \textit{FUM}, we first concatenate texts of user's clicked news as a long user document and transform the user modeling task into a document modeling task.
Then we capture both intra-news and inter-news word-level behavior interactions across the long user document to understand user interest in a fine-grained way.
In addition, the vanilla transformer network is usually inefficient for modeling long documents due to its quadric complexity, thereby we adopt an efficient transformer network named Fastformer~\cite{wu2021fastformer} to capture fine-grained behavior interactions in \textit{FUM}.
We conduct extensive experiments on two real-world datasets.
Experimental results demonstrate that \textit{FUM} can achieve better performance than many news recommendation methods and meanwhile efficiently model user interest.

\section{FUM}

\subsection{Problem Formulation}

We assume that a news article $h$ induces $k$ genres of textual information (e.g., titles and entities) $[T_1,T_2,...,T_k]$, where $T_i$ is the $i$-th genre of the news text.
The textual sequence $T_i$ is composed of multiple tokens: $T_i = [t_{i,1},t_{i,2},...,t_{i,l}]$, where $t_{i,j}$ is the $j$-th word token in the sequence $T_i$, and $l$ is the maximum length of the sequence.
Besides, we assume that a target user $u$ has previously clicked $m$ news, where $h_j$ denotes the $j$-th clicked news.
The news recommendation task aims to mine user interest from user's reading history to further match candidate news for recommendation.
Our work focuses on accurately and efficiently modeling user interest from fine-grained word-level interactions across clicked news.

\subsection{Fine-grained and Fast User Modeling}

The framework of our \textit{FUM} approach is shown in Fig.~\ref{fig.user_model}.
\textit{FUM} is composed of a fine-grained user model and a coarse-grained user model.
The fine-grained user model is used to capture user interest from word-level behavior interactions.
Its core is to concatenate user's clicked news as a long document and capture intra- and inter-news word-level interactions to model user interest.
Specifically, we first encode the $i$-th genre of news text $T_i$ into a text embedding sequence $\textbf{T}_i \in \mathbb{R}^{l\times d}$ via a genre-specific embedding layer, where $d$ is embedding dimension.
Then we concatenate texts sequences of user's reading history into a long sequence $\textbf{T}\in \mathbb{R}^{mkl\times d}$:
\begin{equation}
    \textbf{T} = [\textbf{T}^1_1;...;\textbf{T}^1_k;...;\textbf{T}^m_1;...;\textbf{T}^m_k],
\end{equation}
where $\textbf{T}^i_j$ is the $j$-th text embedding sequence of the $i$-th clicked news $h_i$ and $;$ is the concatenation operation.
Besides, different genres of news texts usually have different semantic characteristics and meanwhile the positional information of texts are also important for semantic understanding.
Thus, to further enrich the embedding sequence of the user document, we concatenate text embeddings of each token with its genre and position embeddings and build a behavior embedding sequence $\textbf{H}\in \mathbb{R}^{L\times g}$, where $g$ is dimension of the concatenated token embedding, and $L$ denotes the total length (i.e., $mkl$) of the behavior embedding sequence.


\begin{table*}[]
\caption{News recommendation performance of different methods on \textit{MIND} and \textit{Feeds}. The improvement of \textit{FUM} over baseline methods is significant at level $p<0.001$ based on t-test.}

\centering
\resizebox{0.95\textwidth}{!}{
\begin{tabular}{ccccc|cccc}
\Xhline{1.5pt}
         & \multicolumn{4}{c|}{\textit{MIND}}                                          & \multicolumn{4}{c}{\textit{Feeds}}                                         \\ \hline
         & AUC            & MRR            & nDCG@5          & nDCG@10        & AUC            & MRR            & nDCG@5         & nDCG@10        \\ \hline
\textit{GRU}      &65.47$\pm$0.18 &31.15$\pm$0.22 &33.64$\pm$0.24 &39.34$\pm$0.24  &62.95$\pm$0.13 &27.57$\pm$0.08 &31.55$\pm$0.12 &37.18$\pm$0.11 \\
\textit{DKN}      &67.19$\pm$0.13 &32.97$\pm$0.19 &35.87$\pm$0.22 &41.53$\pm$0.17  &64.02$\pm$0.25 &28.65$\pm$0.13 &32.97$\pm$0.17 &38.54$\pm$0.17 \\
\textit{NPA}      &67.42$\pm$0.15 &32.97$\pm$0.18 &35.90$\pm$0.23 &41.54$\pm$0.20  &64.83$\pm$0.47 &29.21$\pm$0.36 &33.64$\pm$0.47 &39.18$\pm$0.48 \\
\textit{KRED}     &67.77$\pm$0.15 &33.39$\pm$0.15 &36.34$\pm$0.17 &42.04$\pm$0.15  &64.92$\pm$0.14 &29.27$\pm$0.08 &33.71$\pm$0.13 &39.25$\pm$0.12 \\
\textit{GNewsRec} &68.38$\pm$0.09 &33.46$\pm$0.22 &36.44$\pm$0.23 &42.15$\pm$0.20  &65.02$\pm$0.11 &29.28$\pm$0.10 &33.74$\pm$0.13 &39.28$\pm$0.13 \\
\textit{NAML}     &68.16$\pm$0.11 &33.31$\pm$0.07 &36.26$\pm$0.10 &41.94$\pm$0.08  &65.31$\pm$0.12 &29.47$\pm$0.07 &33.99$\pm$0.09 &39.57$\pm$0.12 \\
\textit{NRMS}     &68.33$\pm$0.27 &33.55$\pm$0.27 &36.53$\pm$0.32 &42.18$\pm$0.30  &65.21$\pm$0.12 &29.39$\pm$0.05 &33.87$\pm$0.06 &39.46$\pm$0.08 \\
\textit{LSTUR}    &68.53$\pm$0.10 &33.58$\pm$0.15 &36.54$\pm$0.18 &42.23$\pm$0.17  &65.31$\pm$0.20 &29.54$\pm$0.15 &34.08$\pm$0.19 &39.63$\pm$0.19 \\
\textit{FIM}      &68.15$\pm$0.33 &33.36$\pm$0.27 &36.38$\pm$0.30 &42.02$\pm$0.31  &65.47$\pm$0.12 &29.62$\pm$0.07 &34.19$\pm$0.09 &39.72$\pm$0.09 \\
\hline
\textit{FUM}       & \textbf{70.01}$\pm$0.10 & \textbf{34.51}$\pm$0.13 & \textbf{37.68}$\pm$0.14 & \textbf{43.38}$\pm$0.13 &\textbf{66.93}$\pm$0.19 &\textbf{30.49}$\pm$0.16 &\textbf{35.31}$\pm$0.21 &\textbf{40.87}$\pm$0.18 \\ 
\Xhline{1.5pt}
\end{tabular}
}
\label{table.pe}
\end{table*}

The transformer network~\cite{vaswani2017attention} is an effective technique for document modeling.
However, due to its quadratic complexity, the vanilla transformer network cannot efficiently model long documents.
Fortunately, some efficient transformer methods have been proposed.
To model fine-grained behavior interactions across the long user document, we employ a SOTA efficient transformer network named \textit{Fastformer}~\cite{wu2021fastformer}.
Take an arbitrary attention head as example, the core idea of \textit{Fastformer} is to first summarize global contexts into an embedding $\textbf{q}$ and then transform embeddings of each token based on their relatedness with global contexts:
\begin{equation}
    \textbf{q} = Att(\textbf{q}_1,...,\textbf{q}_L),\quad \textbf{q}_i = \textbf{W}_q \textbf{h}_i,
\end{equation}
\begin{equation}
    \textbf{k} = Att(\textbf{q}*\textbf{k}_1,...,\textbf{q}*\textbf{k}_L),\quad \textbf{k}_i = \textbf{W}_k \textbf{h}_i
\end{equation}
\begin{equation}
    \hat{\textbf{h}}_i = \textbf{W}_o(\textbf{k} * \textbf{v}_i), \quad  \textbf{v}_i=\textbf{W}_v\textbf{h}_i
\end{equation}
where $\textbf{h}_i$ and $\hat{\textbf{h}}_i$ denote the input and output of the $i$-th token in the behavior embedding sequence, $*$ denotes element-wise product, $Att(\cdot)$ denotes the attention pooling network and $\textbf{W}_q$, $\textbf{W}_k$, $\textbf{W}_v$ and $\textbf{W}_o$ denote trainable projection parameters. We remark that \textit{Fastformer} can also be replaced by other efficient transformers.
Then we can concatenate outputs of different attention heads and build a unified contextual representations $\textbf{g}_i$ for the $i$-th token. 
Next, we adopt an attention network to learn embeddings for each clicked news by aggregating embeddings of their tokens:
\begin{equation}
    \textbf{f}_i = Att(\textbf{g}_{(i-1)kl+1},\textbf{g}_{(i-1)kl+2},...,\textbf{g}_{ikl}),
\end{equation}
where $\textbf{f}_i$ represents the $i$-th clicked news.
Finally, we pooling them to build the user embedding $\textbf{u}^f = Att(\textbf{f}_1,...,\textbf{f}_m)$.
In this way, we can efficiently and effectively model and encode user interest from word-level fine-grained behavior interactions.


Besides, we also adopt a coarse-grained user model to better summarize user interest from news-level behavior interactions.
We first apply a news encoder to transform user's clicked news into embeddings.
Details of the news encoder is introduced in Sec.~\ref{sec.ne}.
Then we apply a transformer network to model news-level behavior interactions across user's clicked news, where $\textbf{c}_i$ is the contextualized embedding of $h_i$.
Finally, we build a user embedding $\textbf{u}^c = Att(\textbf{c}_1,...,\textbf{c}_m)$ from news-level behavior interactions and aggregate it with $\textbf{u}^f$ to form a unified user embedding $\textbf{u} = \textbf{u}^f + \textbf{u}^c$.

\subsection{News Encoder}
\label{sec.ne}
Next, we briefly introduce the architecture of the news encoder in \textit{FUM}.
For the $i$-th genre of news text, we apply a text encoder to learn a genre-specific news embedding $\textbf{t}_i$ from $\textbf{T}_i$.
Motivated by \citet{ge2020graph}, the text encoder is implemented by the stack of a transformer and an attention network.
Then, we attentively aggregate genre-specific news embeddings to learn the news embedding $\textbf{n}$.

\subsection{News Recommendation}

Following \citet{wu2019neurald,wu2021uag}, we match the target user $u$ and the candidate news $n$ based on the inner product of their embeddings $r = \textbf{u}^T \textbf{n}$.
Then candidate news are ranked based on their matching scores $r$ for news recommendation.
Besides, we train models based on the BPR loss~\cite{rendle2009bpr}: $\mathcal{L} = -\frac{1}{|\mathcal{D}|} \sum_{i=1}^{|\mathcal{D}|} \sigma({r^c_i-r^n_i}),$
where $\mathcal{D}$ is the training data set, $\sigma$ is the sigmoid function, $r^p_i$ and $r^n_i$ are matching scores for the $i$-th clicked and non-clicked news sample.

\section{Experiment}

\subsection{Dataset and Experimental Settings}

We conduct experiments on two real-world datasets: \textit{MIND} and \textit{Feeds}.
\textit{MIND} is a public dataset based on user data sampled from Microsoft News~\cite{wu2020mind}.
\textit{Feeds} is based on user data sampled from the news feeds platform of Microsoft during Jan. 23 to Apr. 01, 2020 (13 weeks).
We select 200,000 news impressions in the first ten weeks for training and validation, and 100,000 impressions in the last three weeks for evaluation.
Codes are in \url{https://github.com/taoqi98/FUM}.


In experiments, we utilize news topic labels, description texts of entities, titles, and abstracts for news modeling.
Their embeddings are initialized by 300-dimensional glove embeddings~\cite{pennington2014glove} and fine-tuned in experiments.
Besides, we adopt users' recent 50 clicked news to model interest.
In \textit{FUM}, the transformer and \textit{Fastformer} networks are set to $20$ heads, and each head outputs $20$-dimensional vectors.
The attention networks are implemented by MLP networks.
We adopt Adam~\cite{kingma2014adam} with $0.0001$ learning rate to train models for $2$ epoch.
We tune hyper-parameters based on the validation set.



\begin{table*}[]
\caption{Efficiency comparison of user modeling methods on both model training and inference based on 1k samples.}

\centering
\resizebox{0.92\textwidth}{!}{
\begin{tabular}{ccccccccccc} \Xhline{1.5pt}
               & GRU    & DKN    & NAML  & NPA    & KRED   & GNewsRec & LSTUR  & NRMS   & FIM     & FUM    \\ \hline
Training Time  & 11.46s  & 8.19s & 7.98s & 8.10s  & 10.40s & 10.72s   & 11.53s & 11.39s & 15.85s  & 13.21s \\
Inference Time & 2.41s & 44.90s & 1.23s & 1.15s & 1.24s  & 86.90s   & 2.43s  & 2.16s  & 350.38s & 2.75s \\
Cacheable          & \Checkmark & \XSolidBrush & \Checkmark & \Checkmark & \Checkmark & \XSolidBrush & \Checkmark & \Checkmark & \XSolidBrush & \Checkmark \\ \Xhline{1.5pt}
\end{tabular}
}
\label{table.eff}
\end{table*}

\subsection{Performance Evaluation}

We compare \textit{FUM} with several SOTA news recommendation methods:
(1) \textit{GRU}~\cite{okura2017embedding}: propose to build user embeddings via a GRU network.
(2) \textit{DKN}~\cite{wang2018dkn}: propose an attentive memory network to learn user embeddings.
(3) \textit{NPA}~\cite{wu2019npa}: propose a personalized attention mechanism to learn news and user embeddings.
(4) \textit{KRED}~\cite{liu2020kred}: propose a knowledge-aware graph network to learn news embeddings from news titles and entities.
(5) \textit{GNewsRec}~\cite{hu2020graph}: model user interest from user-news graph via a GRU and GNN network.
(6) \textit{NAML}~\cite{wu2019ijcai}: learn user embeddings via an attention network.
(7) \textit{LSTUR}~\cite{an2019neural}: model long- and short-term user interest via GRU network and user IDs.
(8) \textit{NRMS}~\cite{wu2019neuralc}: learn user embeddings via self-attention networks.
(9) \textit{FIM}~\cite{wang2020fine}: model user interest in news from the matching of news texts and reading history via CNN network.

\begin{figure}
    \centering
    \resizebox{0.50\textwidth}{!}{
    \includegraphics{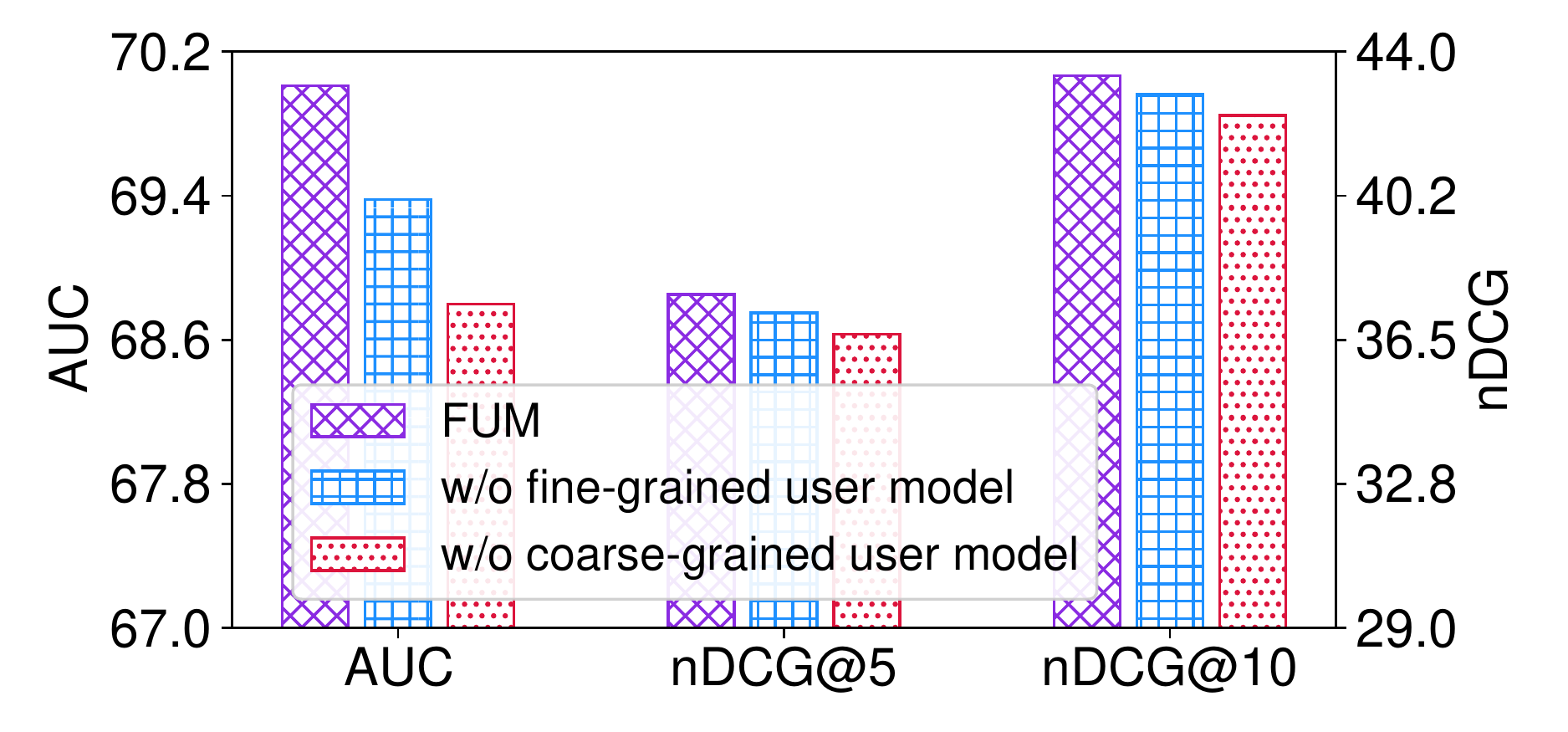}
    }
    \caption{Ablation study of our \textit{FUM} approach.}
    \label{fig.ablation}
\end{figure}

We repeat experiments of different methods 5 times and show average results and standard deviations in Table~\ref{table.pe}.
Results show that \textit{FUM} can achieve much better performance than baseline methods, e.g., \textit{LSTUR} and \textit{NRMS}.
This is because baseline methods can only capture news-level behavior interactions to model user interest.
This is because fine-grained behavior interactions across user's clicked news at word-level contain rich detailed clues to understand user interest.
However, baseline methods usually neglect the fine-grained behavior interactions and thereby only achieve inferior performance. 
Different from these methods, in \textit{FUM} we concatenate texts of user's reading history as a long document and apply an efficient transformer network to capture the fine-grained behavior interactions.
Thus, our \textit{FUM} approach can more accurately model user interest from fine-grained behavior interactions and achieve more effective news recommendation performance.

\subsection{Efficiency Comparison}

Next, we compare the efficiency of different methods on both model training and inference.
In Table~\ref{table.eff}, we first summarize the average time of different methods for training and inferring $1000$ samples.
Due to the space limitation, we only show results on \textit{MIND} in the following sections.
According to Table~\ref{table.eff}, \textit{FUM} achieves comparable or better efficiency than methods that neglects fine-grained behavior interactions.
This is because in \textit{FUM} we adopt a SOTA transformer network proposed for efficient long document modeling to capture fine-grained behavior interactions.
Thus \textit{FUM} can efficiently model fine-grained interactions of the long user document to mine user interest.
Besides, real-world systems usually have strict online latency constraints~\cite{wu2021empowering}.
Thus, in the practice on real-world systems, news and user representations are expected to be offline computed and cached in the platform to improve online efficiency.
Like some baseline methods, news and user representations of \textit{FUM} are also cacheable, which further verify the feasibility of \textit{FUM} in practice.

\begin{figure}
    \centering
    \resizebox{0.50\textwidth}{!}{
    \includegraphics{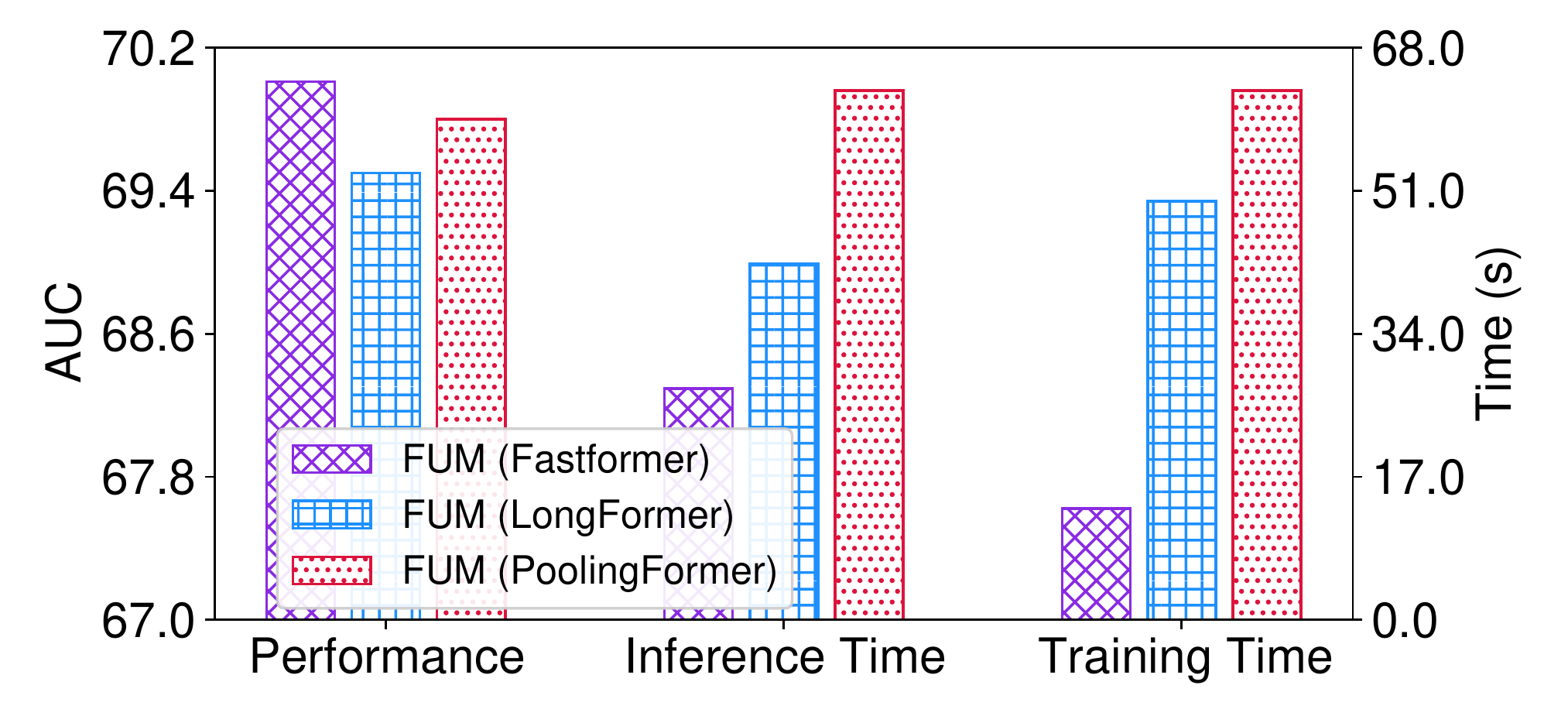}
    }
    \caption{ \textit{FUM} with different efficient transformers. The training and inference time are based on 1k and 10k samples.}
    \label{fig.trans}
\end{figure}

\subsection{Ablation Study}
Next, we conduct an ablation study to verify the effectiveness of the fine- and coarse-grained user model in \textit{FUM} (Fig.~\ref{fig.ablation}).
First, after removing the fine-grained user model, the performance of \textit{FUM} seriously declines.
This is because fine-grained interactions across different clicked news from the same user usually contain rich clues to understand user interest.
The fine-grained user model can effectively capture word-level interactions and better model user interest.
Second, removing the coarse-grained user model also hurts performance.
This is because intra-news behavior interactions are also important for user modeling, which can be effectively captured by the coarse-grained user model in \textit{FUM}.
Besides, the coarse-grained user model also outperforms the fine-grained user model, which may be because intra-news interactions cannot be effectively exploited by the fine-grained model.

\subsection{FUM with Different Efficient Transformers}

Next, we apply different efficient transformers to \textit{FUM} to verify their impacts.
Besides \textit{FastFormer}, we apply two other SOTA efficient transformers, i.e., \textit{LongFormer}~\cite{beltagy2020longformer} and \textit{PoolingFormer}~\cite{zhang2021poolingformer} in \textit{FUM} (Fig.~\ref{fig.trans}).
We first find \textit{FUM} with various transformers can consistently outperform baselines, which verifies the importance of fine-grained user modeling.
Second, \textit{Fastformer} significantly improves efficiency of \textit{FUM} than other transformers.
Thus, we choose \textit{FastFormer} for the fine-grained user modeling in \textit{FUM}.
\section{Conclusion}

In this paper, we propose a fine-grained and fast user modeling framework for news recommendation (named \textit{FUM}), which can understand user interest from fine-grained behavior interactions.
In \textit{FUM}, we first concatenate user's clicked news as a long document.
Then we employ an efficient transformer network named \textit{Fastformer} to capture fine-grained behavior interactions from word-level to target user interest more accurately.
Extensive experiments on two-real world datasets verify that \textit{FUM} can outperform many news recommendation methods and meanwhile efficiently model user interest from fine-grained behavior interactions.




\section*{Acknowledgments}
This work was supported by the National Key Research and Development Project of China under Grant 2018YFB2101501 and Tsinghua-Toyota Joint Research Funds 20213930033.

\bibliographystyle{ACM-Reference-Format}
\bibliography{mybib}


\end{document}